\documentclass{jfm}
\usepackage{natbib}
\usepackage{epsfig}
\usepackage{amssymb}
\usepackage{amsmath}

\def\begineq{\begin{equation}}
\def\endeq{\end{equation}}
\def\be{\begin{equation}}
\def\ee{\end{equation}}

\newcommand{\bm}[1]{\mbox{\boldmath $#1$}}

\title[Fine-scale statistics in convective turbulence]
{Fine-scale statistics of temperature and its derivatives in convective turbulence}
\author[M. S. Emran \& J. Schumacher]%
{M.\ns S.\ns E\ls M\ls R\ls A\ls N \and 
 J.\ns S\ls C\ls H\ls U\ls M\ls A\ls C\ls H\ls E\ls R}
\affiliation{
Department of Mechanical Engineering, Technische Universit\"at Ilmenau,
D-98684 Ilmenau, Germany}
\pubyear{2007}
\volume{999}
\pagerange{1--10}
\date{?? and in revised form ??}
\begin{document}
\maketitle
\begin{abstract}
We study the fine-scale statistics of temperature and  its derivatives in turbulent
Rayleigh-B\'{e}nard convection. Direct numerical simulations are carried out in a cylindrical 
cell with unit aspect ratio filled with a fluid with Prandtl number equal to 0.7 for Rayleigh 
numbers between $10^7$ and $10^9$. The probability density function of the temperature or 
its fluctuations is found to be always non-Gaussian. The asymmetry and  strength of  
deviations from the Gaussian distribution are quantified as a function of the cell height. The deviations of the temperature fluctuations from the local isotropy, as  measured by the skewness of the vertical 
derivative of the temperature fluctuations, decrease in the bulk, but increase in the thermal 
boundary layer for growing Rayleigh number, respectively. Similar to the passive scalar mixing, 
the probability density function of the thermal dissipation rate deviates significantly from a 
log-normal distribution. The distribution is fitted well by a stretched exponential form. The tails become more extended with increasing Rayleigh number which displays an 
increasing degree of small-scale intermittency of the thermal dissipation field for both the bulk and 
the thermal boundary layer. We find that the thermal dissipation rate due to the temperature 
fluctuations is not only dominant in the bulk of the convection cell, but also yields a significant 
contribution to the total thermal dissipation in the thermal boundary layer. This is in contrast
to the ansatz used in scaling theories and can explain the differences in the scaling of the total 
thermal dissipation rate with respect to the Rayleigh number. 
\end{abstract}

\section{Introduction}
Convective turbulence which is driven by the action of buoyancy forces appears in a wide 
range of geophysical and astrophysical systems as well as in numerous technological 
applications such as chip cooling devices or indoor ventilation systems (Kadanoff 2001). 
The main focus of convection experiments has been on the precise quantification of the 
global turbulent heat transport through the cell which is measured by
the dimensionless Nusselt number $Nu$ as a function of the applied outer temperature 
difference, the properties of the working fluid and the geometry (see e.g.
Niemela \& Sreenivasan 2003; Funfschilling {\it et al.} 2005). The three 
dependencies are quantified by dimensionless outer parameters, the Rayleigh number $Ra$, 
the Prandtl number $Pr$ and the aspect ratio $\Gamma$. The numbers are defined as 
\begin{equation}
Ra=\frac{\alpha g \Delta T H^3}{\nu \kappa}\,,\;\;\;\;\;\;\;
Pr=\frac{\nu}{\kappa}\,,\;\;\;\;\;\;\;
\Gamma=\frac{D}{H}\,, 
\end{equation}
where $\alpha$ is the thermal expansion coefficient, $g$ the gravitational acceleration, 
$\Delta T$ the outer temperature difference, $H$ the height of the cell and $D$
its diameter. The Prandtl number compares the kinematic viscosity of the fluid $\nu$ to 
the thermal diffusivity $\kappa$ of the temperature field. The experiments were able to 
explore a large range of Rayleigh numbers in the so-called hard turbulence regime from 
$Ra\sim 10^7$ up to $Ra\sim 10^{17}$ in the case of liquid helium (Niemela {\it et al.} 2000). 
However, most experiments can provide pointwise measurements of time 
series of the turbulent fields only (see e.g. du Puits {\it et al.} 2007 and references therein). 
Only recently, two-dimensional cuts through the flow field have been analysed by combining 
particle image velocimetry and shadowgraph techniques (Xi, Lam \& Xia 2004).     

In contrast, direct numerical simulations (DNS) are currently unable to reach 
in the very high Rayleigh number regime, in particular for aspect ratios $\Gamma\ge 1$. 
The advantage of DNS is however the fully resolved spatial and temporal information on 
the turbulent fields and  local mechanisms of heat transfer (Kerr 1996; 
Verzicco \& Camussi 2003; Hartlep, Tilgner \& Busse 2005; Shishkina \& Wagner 2006;
Shishkina \& Wagner 2007). 

Directly related to the Nusselt number is the mean of the thermal dissipation rate which 
is given by (e.g. Grossmann \& Lohse 2000)
\begin{equation}
\langle\epsilon_{T}\rangle_V=\frac{Nu}{\sqrt{Ra Pr}}\,.
\label{Nueps}
\end{equation}
The thermal dissipation rate field itself measures the magnitude of the temperature gradient  
and is defined as 
\begin{equation}
\epsilon_{T}({\bf x},t)=\kappa \left(\frac{\partial T}{\partial x_i}\right)^2\,. 
\label{epsilont}
\end{equation}      
Here, ${\bf x}=(x,y,z)$ is the three-dimensional position vector and $x_i=x,y$ or $z$.
Both equations, (\ref{Nueps}) and (\ref{epsilont}), imply that the statistics of the fluctuating 
thermal dissipation field is  connected with the local fluctuations of the conductive heat transfer, 
as given by the local currents, ${\bf j}({\bf x},t)=-\kappa ({\bm \nabla}T)$. The measurement 
of the thermal dissipation field --and thus of spatial derivatives-- is experimentally challenging, 
especially in high-Rayleigh number turbulence. Experiments can usually provide well-resolved time 
derivatives (Belmonte \& Libchaber 1996) or temporal increments of the temperature 
field (Zhou \& Xia 2002). In case of temporal increments in convection, closed forms of the corresponding 
probability density functions (PDF) of stretched exponential type have been constructed successfully 
by Ching (1991, 1993). For other flows, such as axisymmetry jets, turbulent channel flows or homogeneous 
isotropic turbulence, PDFs of the velocity increments have been constructed as a superposition of Gaussian distributions 
(Castaing {\it et al.} 1990) or  a product of a Gaussian random variable and a scale-dependent 
random multiplier (Chevillard {\it et al.} 2006). In Rayleigh-B\'{e}nard convection, time derivatives
 however cannot be translated into spatial derivatives by a Taylor frozen-flow hypothesis as in 
pipe or channel flows. Furthermore, for turbulence  in a closed vessel, the concept of 
homogeneity is limited to the cell centre only. He, Tong \& Xia (2007) were recently able to measure 
four temperature signals close to each other simultaneously to reconstruct temperature gradients. 
Their analysis disentangled contributions to the total thermal dissipation coming from the bulk and 
boundary layers. Rayleigh numbers $Ra\sim 10^9$ were attained in the experiments, but the resolution 
of the gradients remained limited to scales larger or equal to the thermal boundary layer thickness. 
Direct numerical simulations by Kerr (1996) and more recently by Shishkina \& Wagner (2007) focused 
on the geometric properties of thermal plumes, the structures that carry the heat away from the bottom 
plate. 

A detailed statistical analysis of the spatial derivatives of the temperature and the thermal dissipation field in 
different regions of the convection cell is thus still missing. In particular, knowledge about the strength 
of the fluctuations around the mean temperature gradient and their dependence on the Rayleigh number 
allows for a validation of predictions by scaling theories, e.g. those by Grossmann \& Lohse (2000). 
Such analysis is also interesting from the perspective of passive scalar mixing in turbulence where 
progress in the understanding of the mechanisms that cause intermittent fluctuations has been made recently 
(Shraiman \& Siggia 2000). A first and open point is to understand the differences between passive and 
active scalars such as the temperature in convective turbulence.  For the passive scalar case, it is 
known that larger amplitudes of the dissipation fields are mostly concentrated on fine scales 
(Kushnir, Schumacher \& Brandt 2006) and that their statistical study puts rather large resolution 
constraints on DNS. This was discussed by Schumacher, Sreenivasan \& Yeung (2005) and Schumacher \& 
Sreenivasan (2005).  

In the present work, we provide a detailed height-dependent statistical analysis of the temperature, its fluctuations and spatial derivatives. Odd-order moments of spatial derivatives along an imposed 
outer mean scalar gradient have been used successfully to quantify deviations from local isotropy in 
shear flows or for the mixing of scalars in turbulence (Warhaft 2002; Pumir 1996; Schumacher \& 
Sreenivasan 2003). We adopt these ideas for the present active scalar case and discuss the dependence of these
anisotropy measures on $Ra$ and in different regions of the convection cell. Finally, we study the 
statistics of the thermal dissipation rate in the bulk and close to the bottom and top plates of the cell.
The results are then related to findings from the passive scalar dissipation field. Our study is intended to 
build a bridge between the mixing of passive and active scalar fields by comparing the statistical 
properties for both cases. The outline of this paper is as follows. In the next section, we present 
the equations of motion and details of the numerical model. The third section discusses the statistics 
of the total temperature, temperature fluctuations and their gradients. The fourth chapter is on the 
statistics and  Rayleigh number dependence of the thermal dissipation field. Concluding 
remarks are given at the end.
  
\section{Equations of motion and numerical model}
The equations for an incompressible three-dimensional Navier--Stokes fluid in the 
Boussinesq approximation are solved in combination with the advection--diffusion equation
for the temperature field. The system is given by
\begin{eqnarray}
\label{nseq}
\frac{\partial{\bf u}}{\partial t}+({\bf u}\cdot{\bf \nabla}){\bf u}
&=&-{\bf \nabla} p+\nu {\bf \nabla}^2{\bf u}+\alpha g T {\bf e}_z\,,\\
\label{ceq}
{\bf \nabla}\cdot{\bf u}&=&0\,,\\
\frac{\partial T}{\partial t}+({\bf u}\cdot{\bf \nabla}) T
&=&\kappa {\bf \nabla}^2 T\,,
\label{pseq}
\end{eqnarray}
where $p({\bf x},t)$ is the pressure, ${\bf u}({\bf x},t)$ the velocity field  and 
$T({\bf x},t)$ the total temperature field. The latter can be decomposed into a mean profile 
$\langle T\rangle_A$ and fluctuations $\theta$,
\begin{equation}
T({\bf x},t)=\langle T\rangle_A(z)+\theta({\bf x},t)\,.
\end{equation}
Two different types of statistical averages are used. The average $\langle\cdot\rangle_V$ is 
taken over the whole cell volume {\em and} a sequence of statistically independent snapshots. 
The average  $\langle\cdot\rangle_A$ is calculated over the circular plane at a fixed height 
$z$ {\em and} a sequence of snapshots. We thus combine always spatial and temporal averages. 
All quantities are expressed in characteristic units. Length scales are  normalized with 
respect to the height of the cell, $H$, velocities with respect to the free-fall velocity, 
$U=\sqrt{\alpha g\Delta T H}$, time scales with respect to $H/U$. Temperatures are given by 
$0\le \tilde{T}=(T-T_c)/\Delta T\le 1$ where $T_c$ is the (cold) temperature at the top plate.  
When measured in these units, the thermal diffusivity becomes $\kappa=1/\sqrt{Ra Pr}$ and the 
kinematic viscosity $\nu=\sqrt{Pr/Ra}$. Our studies are conducted for $Pr=0.7$ in a 
cylindrical container of aspect ratio $\Gamma=1$. The top and bottom plates have no-slip 
boundary conditions, ${\bf u}\equiv 0$, at a fixed temperature. The side walls are adiabatic
no-slip boundaries, i.e., ${\bf u}\equiv 0$ and $\partial T/\partial r=0$.

The equations are discretized on a staggered mesh and solved by a second-order finite 
difference scheme (Verzicco \& Orlandi 1996; Verzicco \& Camussi 2003). The pressure field $p$ 
is determined by a two-dimensional Poisson solver after applying one-dimensional fast Fourier 
transformations in the azimuthal direction. The time advancement is done by a third-order Runge-Kutta 
scheme. The grid spacings are non-equidistant in the radial and axial directions. In the vertical 
direction, they correspond to Tschebycheff collocation points. Table 1 lists the parameters of the 
present simulations: $Ra$, $Pr$, $\Gamma$, the resulting Reynolds and Nusselt numbers and the 
corresponding grid resolutions. The table also shows the grid spacing due to the resolution 
criterion of Gr\"otzbach (1983) (see also Kerr (1996) for a discussion) which is given by
\begin{equation}
\Delta_G\le \pi\eta_K=\pi \left(\frac{Pr^2}{(Nu-1)Ra}\right)^{1/4}
\label{gridspaceG}
\end{equation}    
for $Pr\le 1$. The maximum of the geometric 
mean of the grid spacings in our runs is also provided by 
\begin{equation}
\tilde{\Delta}=\max_{{\bf x}=(\phi,r,z)}\sqrt[3]{\Delta_{\phi}\Delta_r\Delta_z}\,.
\label{gridspace}
\end{equation}    
\begin{table} 
\begin{center}
\begin{tabular}{lcccccccccccc}
No. & $Ra$ & $Pr$ &$\Gamma$ & $Nu$ & $Re$ & $N_{\phi}$ & $N_r$ & $N_z$ &  $\Delta_G$ & $\tilde\Delta$ & $\delta_T$ \\
\hline
1 & $1\times 10^7$ & 0.7 & 1 & 16.7 & 682 & 193 &  97 & 128  & 0.0236  & 0.0104 & 0.0300\\
2 & $5\times 10^7$ & 0.7 & 1 & 25.9 & 1444 & 257 & 129 & 160  & 0.0139  & 0.0080 & 0.0193\\
3 & $1\times 10^8$ & 0.7 & 1 & 31.7 & 2036 & 271 & 151 & 200  & 0.0111  & 0.0069 & 0.0158\\
4 & $5\times 10^8$ & 0.7 & 1 & 52.4 & 4530 & 301 & 201 & 256  & 0.0065  & 0.0058 & 0.0095\\
5 & $1\times 10^9$ & 0.7 & 1 & 65.0 & 6255 & 361 & 181 & 310  & 0.0053  & 0.0051 & 0.0077\\
\end{tabular}
\caption{List of the dimensionless parameters $Ra$, $Pr$ and $\Gamma$ for different runs.
The Reynolds number is $Re=\frac{U_{rms}H}{\nu}$ and the Nusselt number is given by (\ref{Nuz}) and (\ref{Nuz1}). 
$U_{rms}$ is an average over the whole cell volume and a sequence of statistically independent snapshots.
All three velocity components, $u_r$, $u_{\phi}$ and $u_z$, are included. The Reynolds number defined in this way follows $Re=0.33Ra^{0.475}$.
In addition, we list the corresponding grid resolutions in azimuthal, radial and vertical
directions $(N_{\phi}, N_r, N_z)$, the grid spacing due to Gr\"otzbach (1983), $\Delta_G$, 
and the maximum of the geometric mean of the grid spacings for the present simulations as given 
by (\ref{gridspace}), $\tilde\Delta$. $\delta_T$ is the thermal boundary layer thickness, which is 
determined as $H/(2Nu)$.}
\end{center}
\label{tab1}
\end{table}

\begin{figure}
\centerline{\includegraphics[angle=0,scale=0.35,draft=false]{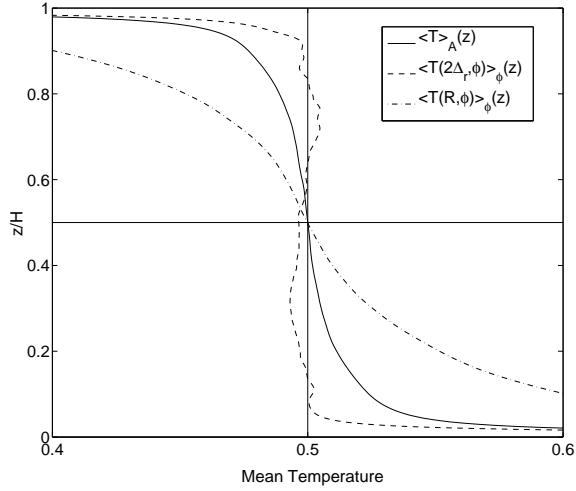}}
\caption{Mean temperature profiles at $Ra=10^8$. The standard profile $\langle T\rangle_A(z)$ is compared with two vertical temperature profiles which are calculated at fixed radial 
positions and averaged  azimuthally. One profile is taken at the sidewall ($r_0=R$), the other profile is close to the centre line ($r_0=2\Delta_r$). All mean profiles vary from 0 to 1 along the horizontal axis. The 
interval $[0.4, 0.6]$ is shown only.}
\label{meanprof}
\end{figure}

Figure \ref{meanprof} shows the vertical mean temperature profiles taken in three different 
ways in the turbulent flow for a Rayleigh number $Ra=10^8$. The standard mean temperature profile 
$\langle T\rangle_A(z)$ is compared with two vertical mean temperature profiles which are conducted at fixed 
radial positions and averaged azimuthally, $\langle T(r_0,\phi)\rangle_{\phi}(z)$. It can be seen that the 
slopes of the temperature vary significantly. While the profile $\langle T\rangle_A(z)$ has almost zero 
slope at $z=0.5$, a destabilizing slope, $d\langle T\rangle_{\phi}(z)/dz <0$, is observed close to the 
sidewall. Close to the centre line, the situation is reversed, we observe a stabilizing slope, 
$d\langle T\rangle_{\phi}(z)/dz >0$. Our finding agrees with a recent experiment by Brown \& Ahlers (2007) 
and confirms that the assumption of the temperature drop $\Delta T$ across the thermal boundary layer is a 
simplification. The presence of a non-vanishing local mean temperature gradient can have an impact on the 
statistics of the small-scale temperature fluctuations, similar to the passive scalar case. This point will
be discussed further in section 3.  

The thermal boundary layer is resolved with at least 14 lateral grid  planes for all 
runs.  
\begin{figure}
\centerline{\includegraphics[angle=0,scale=0.35,draft=false]{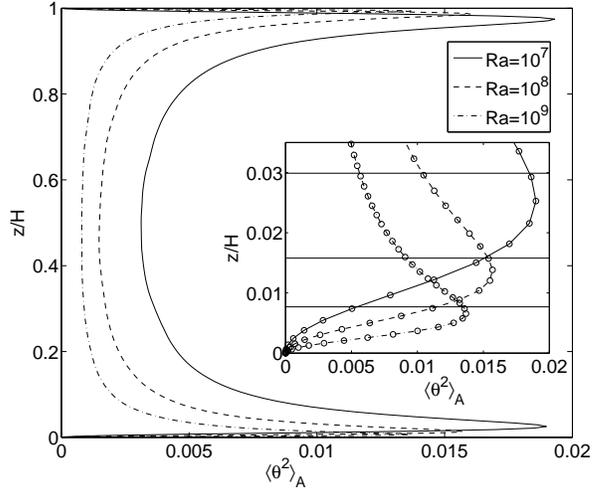}}
\caption{Mean square fluctuations $\langle\theta^2\rangle_A$ as a function of $z/H$ for
Rayleigh numbers $Ra=10^7, 10^8$ and $10^9$ at an aspect ratio $\Gamma=1$. The inset 
magnifies the thermal boundary layer. The solid horizontal lines 
correspond to $\delta_T=H/(2Nu)$.}
\label{thetarms}
\end{figure}
This is indicated in Fig.÷\ref{thetarms} where we plot the vertical mean square 
profiles of the temperature fluctuations and highlight the (non-equidistant) vertical 
grid spacing by the symbols in the inset. The horizontal lines correspond to the thermal boundary 
layer thickness estimated by $\delta_T=H/(2Nu)$, which is always close to the position of 
the maximum of $\theta^2_{rms}(z)=\langle\theta^2\rangle_A(z)$. 
All runs were performed for at least 80 large-scale eddy turnover times in order to 
gather sufficient statistics. We also verified that the vertical Nusselt number profile 
given by
\begin{equation}
Nu(z)=\frac{\langle u_z T\rangle_A-\kappa\frac{\partial\langle T\rangle_A}{\partial z}}
{\kappa\frac{\Delta T}{H}}\,,
\label{Nuz}
\end{equation}
is a constant with a standard deviation of about $1\%$. The mean value 
of the Nusselt number profile, 
\begin{equation}
Nu=\frac{1}{H}\int_0^H Nu(z)\mbox{d}z\,,
\label{Nuz1}
\end{equation}
follows the scaling law $Nu=N_0\times Ra^{\beta}$, with 
$N_0=0.175$ and $\beta=0.283$. 
\begin{figure}
\centerline{\includegraphics[angle=0,scale=0.6,draft=false]{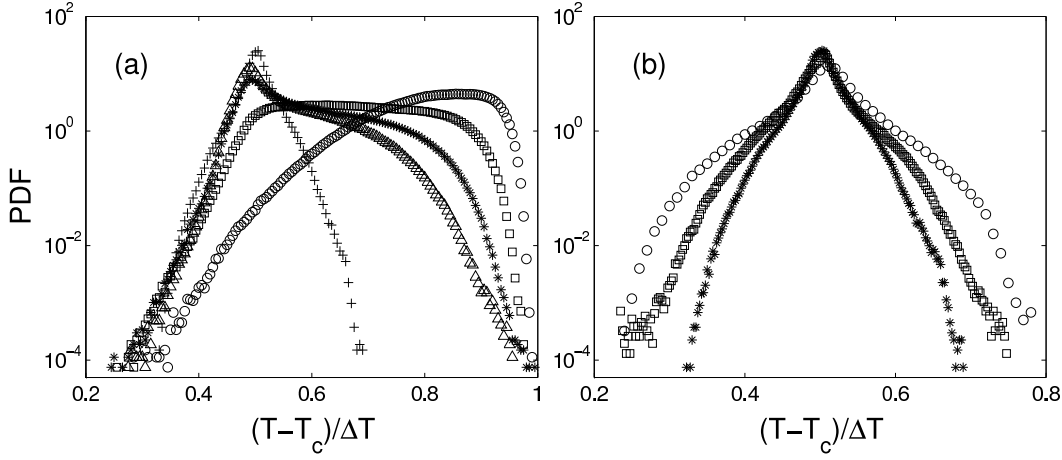}}
\caption{Probability density functions (PDF) of the dimensionless temperature $\tilde{T}$. (a)
All data are taken for $Ra=10^9$ and from the whole plane at height z. Heights are $z=0.5 \delta_T$ ($\circ$), $z=\delta_T$ ($\square$),
$z=2 \delta_T$ ($\ast$), $z=4 \delta_T$ ($\triangle$) and $z=0.5 H$ (+). (b) Data are taken at the 
centre plane $z=0.5$ for different Rayleigh numbers: $Ra=10^7$ ($\circ$), $Ra=10^8$ ($\square$) and 
$Ra=10^9$ ($\ast$).}
\label{PDFtheta_z}
\end{figure}
\begin{figure}
\centerline{\includegraphics[angle=0,scale=0.55,draft=false]{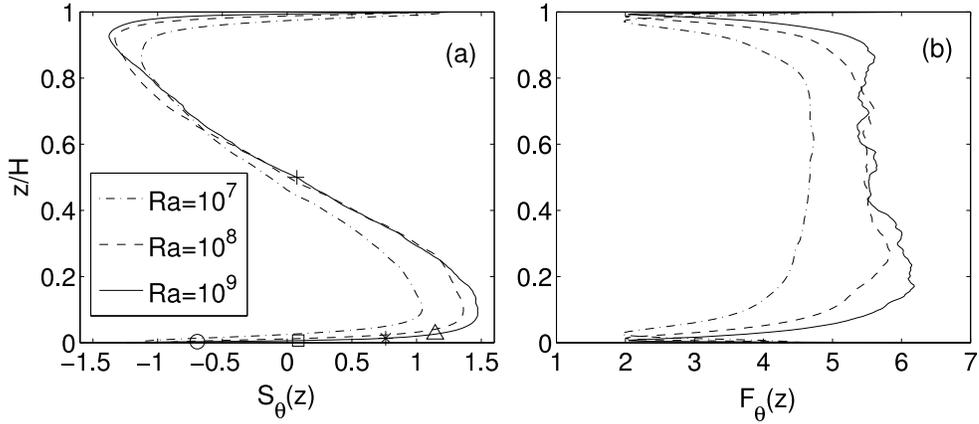}}
\caption{Height dependence of the skewness (left) and flatness (right) of the temperature 
fluctuations. Data are given for three Rayleigh numbers. The symbols in the skewness curve 
of $Ra=10^9$ indicate the distance from the wall at which the PDFs in Fig. \ref{PDFtheta_z}a 
are plotted.}
\label{szfz}
\end{figure}
\begin{figure}
\centerline{\includegraphics[angle=0,scale=0.4,draft=false]{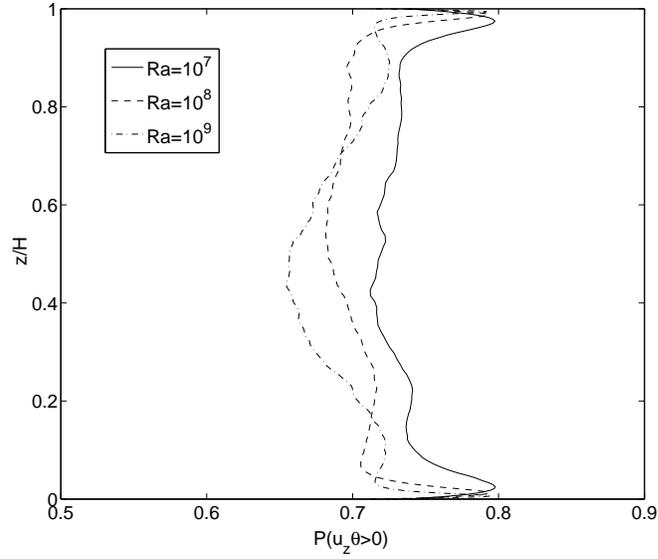}}
\caption{Vertical profiles of the probability of $u_z\theta >0$ for three different 
Rayleigh numbers.}
\label{uztheta}
\end{figure}
\begin{figure}
\centerline{\includegraphics[angle=0,scale=0.5,draft=false]{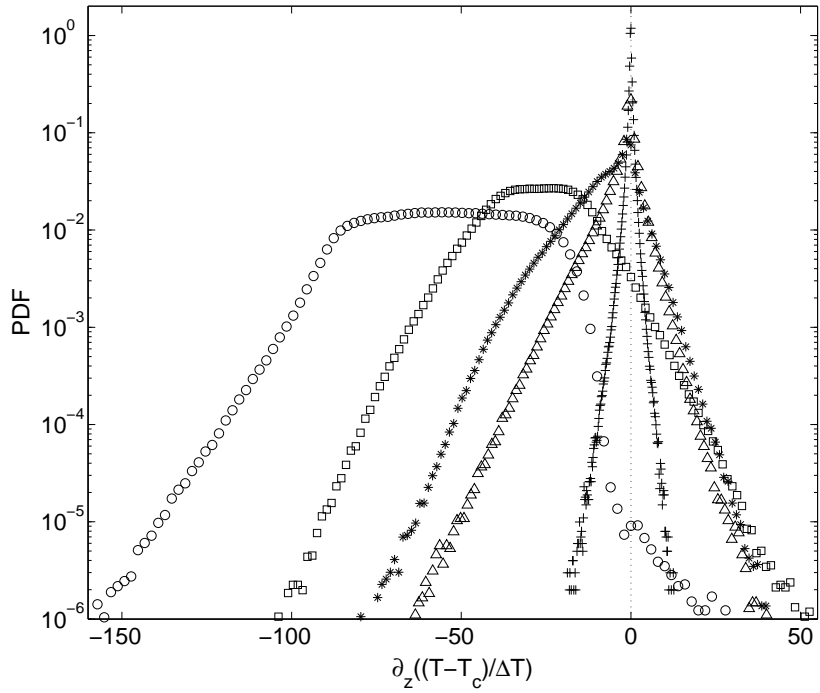}}
\caption{Probability density functions (PDF) of vertical derivative of the temperature, 
$\partial\tilde{T}/\partial z$, taken in the same planes as in Fig. \ref{PDFtheta_z}a. All data are again
for $Ra=10^9$. Heights are $z=0.5 \delta_T$ ($\circ$), $z=\delta_T$ ($\square$), $z=2 \delta_T$ ($\ast$), 
$z=4 \delta_T$ ($\triangle$) and $z=0.5 H$ (+).}
\label{gradpdf}
\end{figure}
\begin{figure}
\centerline{\includegraphics[angle=0,scale=0.55,draft=false]{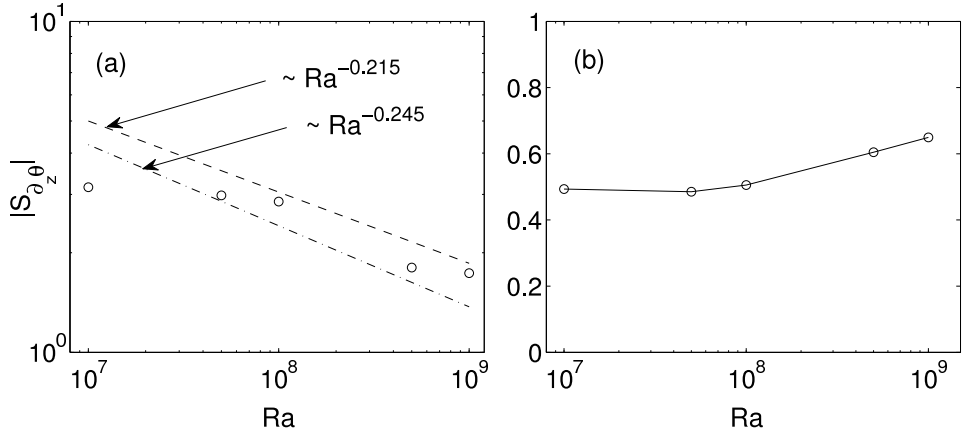}}
\caption{Skewness of the vertical temperature derivative as a function of the Rayleigh number 
$Ra$. (a) Data analysis is performed in the center of the cell for points in 
$V=\{r, \theta, z |\,0.4 H\le z\le 0.6 H\}$. Data are displayed in double-logarithmic axes. 
In addition, the decay laws of the skewness with respect to $Ra$ are shown which would be in line with a return to
isotropy (dashed lines). (b) Analysis is conducted in the thermal boundary layer for points in  
$V=\{r,\theta, z |\,0\le z\le \delta_T\}$. Data are displayed in logarithmic-linear diagram. }
\label{s3grad}
\end{figure}

\section{Temperature and its derivative statistics}
\subsection{Total temperature and its fluctuations}
Figure \ref{PDFtheta_z} shows the probability density functions (PDF) of the dimensionless 
temperature $\tilde{T}$ at different heights from the bottom plate and as a function of the
Rayleigh number, respectively. The findings are in agreement with Siggia (1994) and Kerr 
(1996). As expected, the PDFs in Fig. \ref{PDFtheta_z}a become increasingly symmetric toward 
the centre of the convection cell. This height-dependent asymmetry can be measured alternatively 
by the skewness of the temperature fluctuations, $\theta$, which is given by 
\begin{equation}
S_{\theta}(z)=\frac{\langle\theta^3\rangle_{A}}{\langle\theta^2\rangle_{A}^{3/2}}\,.
\end{equation}
The graphs are shown in Fig. \ref{szfz}a. A small skewness around $\delta_T$ implies that 
rising and falling plumes are observed at the edge of the bottom and top thermal boundary layers. 
Simultaneously, the colder and hotter pockets of temperature are generated in the vicinity of 
the thermal plumes which detach from the thermal boundary layer. This is a consequence of the   
incompressibility of the fluid as investigated more detailed in the Lagrangian frame of reference
by Schumacher (2008). Up to a height of $z/H\approx0.1$, the skewness increases monotonically 
before it declines to zero at the cell centre plane. The plots show that the maximum value of skewness 
increases with increasing Rayleigh number. A large positive value of the skewness close to the 
bottom plate means that rising plumes with $T(z)>\langle T\rangle_A(z)$ are dominant in this 
region of the convection cell. This is exactly the region in which a  cusp is forming at $\tilde{T}=0.5$
in the PDF (see Fig. \ref{PDFtheta_z}a). 

The distribution of the temperature in the centre plane ($z=0.5$) of the cell as a function of the
Rayleigh number is shown in Fig. \ref{PDFtheta_z}b. The profiles are clearly non-Gaussian for all three 
Rayleigh numbers. The support of the PDF becomes narrower with  increasing $Ra$. The PDF for the fluctuations $\theta$ has the same shape in the centre plane. We found that the PDFs of $\theta$ for the two larger Rayleigh numbers collapse when the temperature 
fluctuation is normalized by the corresponding root-mean-square value, $\theta_{rms}$, at $z=0.5$. The PDF
for $Ra=10^7$ has slightly smaller tails. The magnitude of the deviations from Gaussianity can be 
measured by the flatness which is given by
\begin{equation}
F_{\theta}(z)=\frac{\langle\theta^4\rangle_{A}}{\langle\theta^2\rangle_{A}^{2}}\,.
\end{equation}
Figure \ref{szfz}b shows the vertical profiles of the flatness for the three Rayleigh 
numbers. The non-Gaussianity, i.e., a flatness larger than 3, is present for all three $Ra$.
This is in agreement with other studies such as by Heslot {\it et al.} (1987), Castaing {\it et al.} 
(1989) or Ching (1991). It can also be observed that the flatness value levels off in the cell centre
for the two larger Rayleigh numbers. This could indicate that the transition to the so-called hard turbulence regime of
convection has been completed. Dimotakis (2005) discussed a mixing transition for passive scalars at $Re\sim10^4$. Above this threshold, a weaker $Re$ dependence was predicted. A similar behaviour could be the reason for our observation. All flatness profiles have a minimum right above the thermal boundary 
layer. The slight asymmetry of the profile at $Ra=10^9$ is due to limitations in the statistical analysis. 
Even a longer time advancement and 140 statistically independent snapshots were not sufficient to obtain 
a symmetric profile since this analysis of the higher-order moments is done plane by plane.

As already seen, the non-Gaussian PDF of the total temperature and the temperature fluctuations in 
the cell centre is a robust feature in thermal convection. It has been reported in experiments 
starting with Heslot {\it et al.} (1987) and Castaing {\it et al.} 
(1989). An almost exponential shape was emphasized by Yakhot 
(1989) on the basis of a hierarchy of moment equations for the temperature fluctuations. These ideas have been
extended by Ching (1993) later. Yakhot suggested that this form  prevails even for moderate Rayleigh numbers (which 
implies moderate Reynolds numbers), whenever regions with $u_z \theta>0$ are dominant. A correlation 
between the vertical velocity and temperature fluctuation is nothing else but a fingerprint of 
a local blob of heat transferred through the cell. Inspection of our data yields indeed that 
larger regions with $u_z \theta>0$ are present everywhere in the cell, even in the centre where 
coherent plumes are absent. This is shown in Fig. \ref{uztheta} where we plot the vertical 
dependence of the probability $P(u_z\theta>0)$ for three different Rayleigh numbers. It can 
be clearly seen that the probability decreases with increasing Rayleigh number, but remains 
significantly larger than 0.5 for all three cases. The plot is another manifestation of the net
transfer of heat through the cell.

Interestingly, the passive scalar turns out to be more sensitive to a particular driving and 
the Reynolds number of the advecting flow. When a mean scalar gradient is absent, the statistics
is close to Gaussian or sub-Gaussian such as in Mydlarski \& Warhaft (1998) or Watanabe \& Gotoh
(2004). In case of a non-vanishing  mean scalar gradient, Pumir {\it et al.} (1991) suggested an 
exponentially distributed passive scalar as a generic feature, based on a simple one-dimensional 
random advection model. This was confirmed experimentally for sufficiently high Reynolds numbers 
(Gollub {\it et al.} 1991; Jayesh \& Warhaft 1991; Jayesh \& Warhaft 1992). However, sub-Gaussian
distributions of $\theta$ have also been found for this case by Overholt \& Pope (1996), Ferchichi \& Tavoularis 
(2002) and Schumacher \& Sreenivasan (2005). Experiments by Gylfasson \& Warhaft (2004) detected 
that the statistics depends on the particular driving of the turbulent flow. While an active grid 
caused sub-Gaussian fluctuations, a fine static grid caused super-Gaussian passive scalar distributions. 
Furthermore the deviations from Gaussian distributions decreased with increasing downstream distance in their
experiments. The two findings imply that the ratio of the integral scales to the system scale is 
important. This is  a point which has to be addressed more detailed.  

\subsection{Vertical derivative of temperature fluctuations}
An important question in turbulent convection is if the fluctuations of the temperature field in the
bulk can be considered as being locally isotropic. In order to shed light on this point we adopt an
approach that has been used successfully for passive scalars with a mean scalar gradient (Warhaft 2000). 
As we have seen in section 2, the local mean temperature gradient can deviate from zero while the overall 
profile sums up to a mean temperature gradient which is almost zero (see Fig. 1). Following Brown
\& Ahlers (2007), we define a ratio $\Xi$ that relates a difference of azimuthally averaged mean temperatures 
at fixed radial distance $r_0$ to the total temperature drop. It is given by
\begin{equation}
\Xi(r_0)=\frac{\langle T(r_0)\rangle_{\phi}(z=3H/4)-\langle T(r_0)\rangle_{\phi}(z=H/4)}{\Delta T}\,.
\end{equation} 
Close to the side wall, we get $\Xi(R)=-0.12$ for $Ra=10^7$, $\Xi(R)=-0.07$ for $Ra=10^8$ and 
$\Xi(R)=-0.04$ for $Ra=10^9$. This large slope is mainly due to the rising and falling plumes. At the centre line, 
it follows $\Xi(2\Delta_r)=0.015$ for $Ra=10^7$, $\Xi(2\Delta_r)=0.009$ for $Ra=10^8$ and $\Xi(2\Delta_r)=
0.006$ for $Ra=10^9$. Although decreasing with increasing Rayleigh number for both cases, their magnitude is not negligible. We observe that local mean temperature gradients are present. This supports the idea to conduct an analysis similar to passive 
scalar turbulence.  

The statistics of the vertical derivative of the temperature, $\partial\tilde{T}/\partial 
z$, is shown in Fig. \ref{gradpdf}. The PDFs are analysed at the same heights as those for the 
temperature in Fig. \ref{PDFtheta_z}a. A sharp decrease of the PDF at $0.5\delta_T$ is present in the vicinity
of $\partial\tilde{T}/\partial z=0$, simply because positive vertical derivatives are very unprobable close to the bottom
plate. With increasing distance from the plate, the support of the PDF decreases. A negative skewness of all 
distributions can be observed which implies preferential derivatives along the mean negative temperature 
gradient. Recall also that a turbulent field is thought to be perfectly locally isotropic when all odd-order 
derivative moments are exactly zero. Therefore, the derivative skewness with respect to the temperature 
fluctuations $\theta$, defined as
\begin{equation}
S_{\partial_z\theta}(z)=\frac{\big\langle\left(\frac{\partial \theta}{\partial z}\right)^3\big\rangle_{V}}
                             {\big\langle\left(\frac{\partial \theta}{\partial z}\right)^{2}\big
                             \rangle_{V}^{3/2}}\,,
\end{equation} 
is presented here. It
measures deviations from the local isotropy at the smaller scales of convective turbulence. 
Figure \ref{s3grad} shows that such deviations are indeed present in different regions of the 
cylindrical domain.  With increasing Rayleigh number and thus increasing Reynolds number, the 
derivative skewness $S_{\partial_z\theta}(z)$  decreases in magnitude in the 
centre region (see Fig. \ref{s3grad}a). Following the original idea on the return to local 
isotropy in a simple shear flow (Lumley 1967) and its adaption to the passive scalar case
(Warhaft 2000), a return would require a rather rapid decay law of the skewness with respect
to the Taylor microscale Reynolds number $R_{\lambda}$, i.e. $|S_{\partial_z\theta}| \sim R_{\lambda}^{-1}$. This   
was not found in passive scalar case for Schmidt numbers around unity (Warhaft 2000). Given the 
scaling dependence of the Reynolds number with respect to the Rayleigh number, which will be discussed more detailed 
in section 4.1, and the relation $Re\sim R_{\lambda}^2$, a return to local isotropy in convection 
would require a decay law which is given by 
\begin{equation}
|S_{\partial_z\theta}|\sim Ra^{-\zeta/2}\,,
\end{equation}
for a fixed $Pr$. Here $\zeta$ is the scaling exponent for $Re\sim Ra^{\zeta}$ relation, which was found to 
vary between $\zeta=0.43$ and 0.49 in experiments (see section 4.1). Figure \ref{s3grad}a indicates that 
our data follow such a decay law for the larger Rayleigh numbers. This is in line with the observation
that the local mean gradients in the bulk decrease in magnitude with increasing $Ra$. The mean 
temperature gradient still causes ramps and cliffs of the temperature and thus a non-vanishing 
derivative skewness, but its impact seems to be weaker compared to a passive scalar which lacks a
return to isotropy. A larger 
range of Rayleigh numbers and the analysis of the hyperskewness would allow to draw a firm 
conclusion.  

The opposite trend with respect to the 
Rayleigh number can be observed in the thermal boundary layer where the skewness 
$S_{\partial_z\theta}(z)$ is increasing in magnitude with Rayleigh number $Ra$ (see Fig. \ref{s3grad}b). Figure \ref{gradpdf} shows 
that particularly the PDFs in the boundary layer are strongly skewed to negative derivatives.  This 
observation is due to the fact that  the heat transfer is  mostly contributed by the second term of (\ref{Nuz}) there and that the fluctuations about the mean are consequently dominated by the conductive contributions  
$j_z=-\kappa (\partial\theta/\partial z)$. We will come back to this point in the next section when 
discussing the height dependence of the thermal dissipation.  
      
\section{Thermal dissipation rate}
\subsection{Vertical profiles and Rayleigh number scaling}
\begin{figure}
\begin{center}
\includegraphics[angle=0,scale=0.35,draft=false]{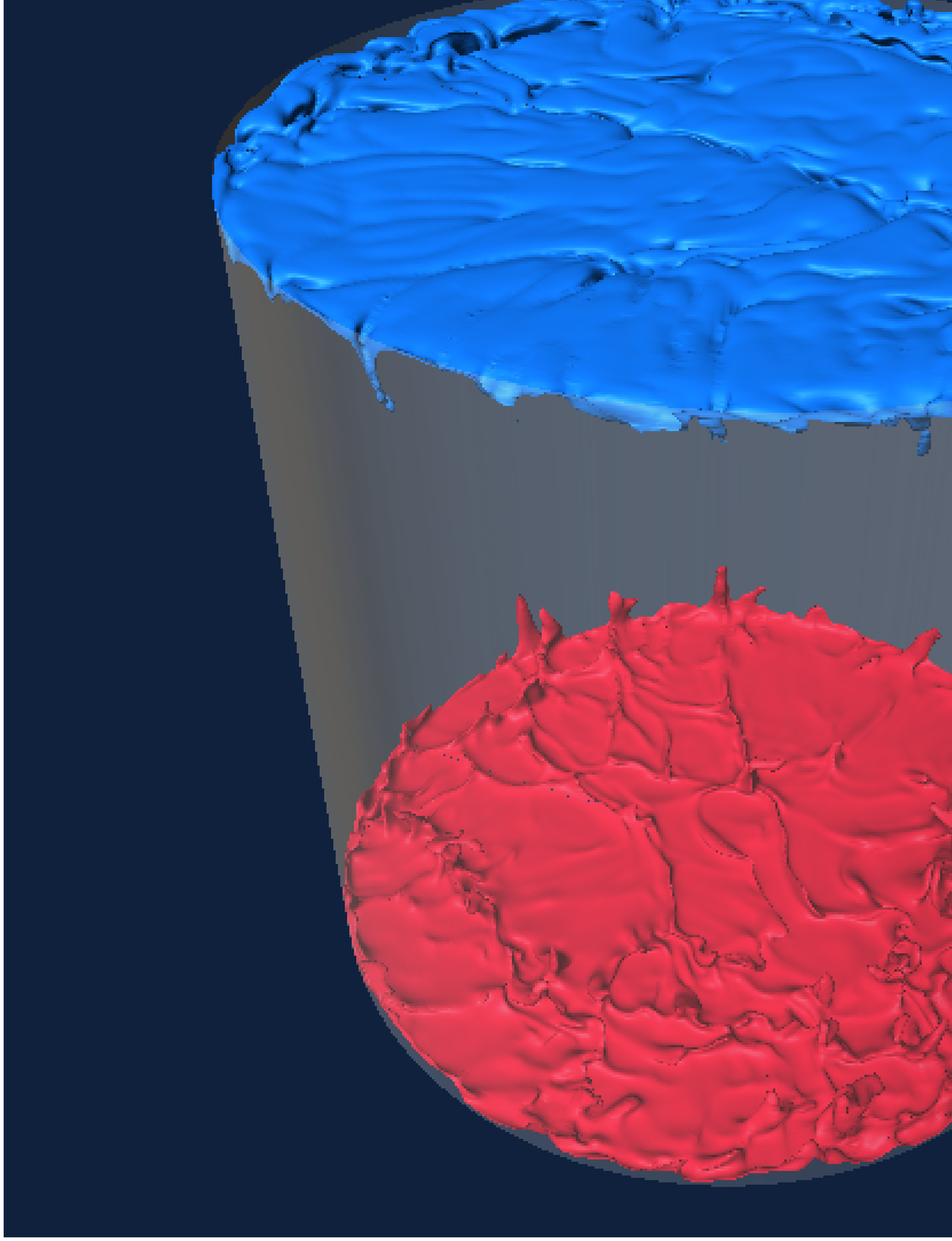}

\vspace{0.5cm}
\includegraphics[angle=0,scale=0.35,draft=false]{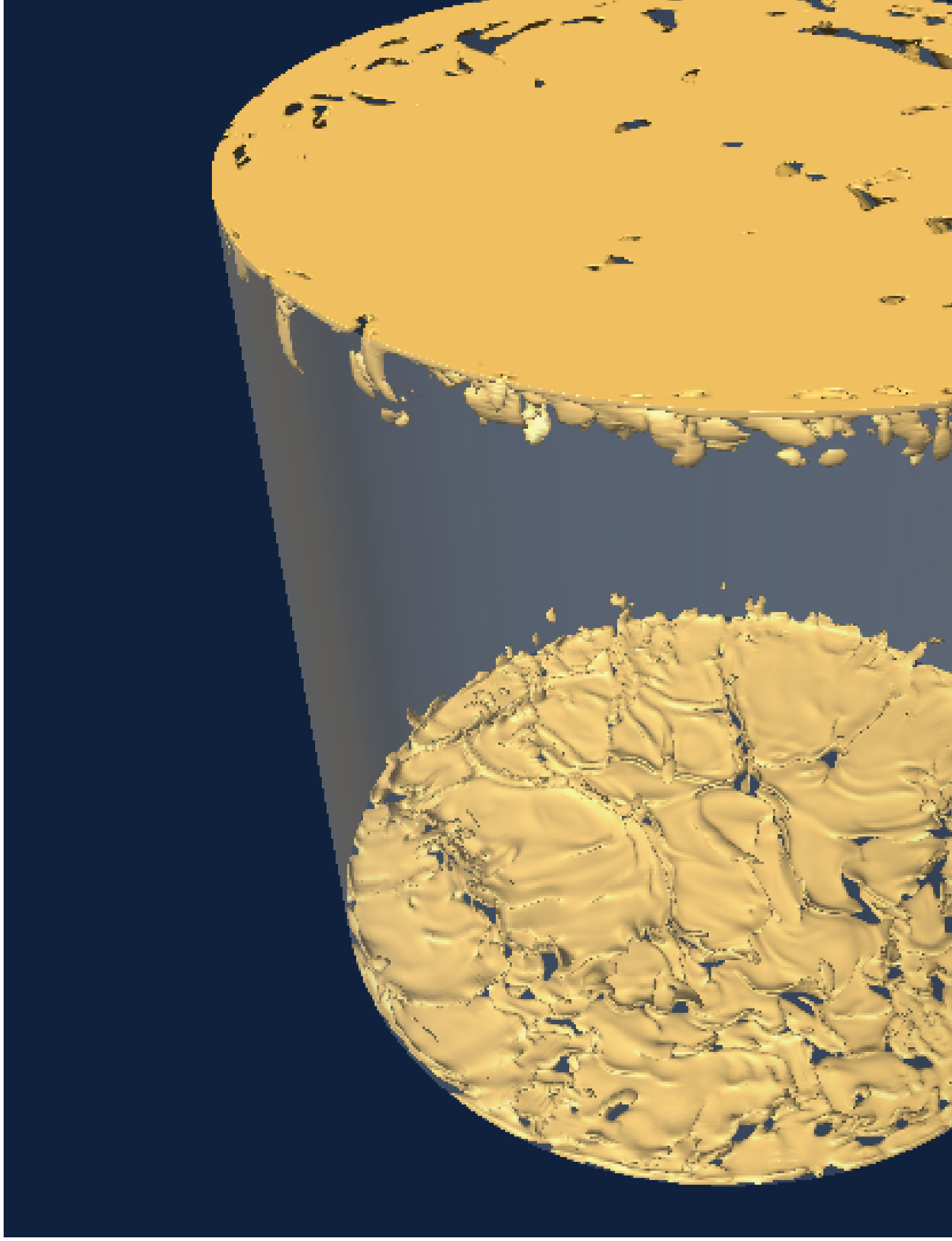}
\caption{Snapshots of the turbulent convection at $Ra=10^9$. Top: Isosurfaces of the temperature $T$
are taken at 0.7 (bottom plate) and 0.3 (top plate), respectively. Bottom: The corresponding thermal 
dissipation field $\epsilon_T$ is shown at the level 10$\langle\epsilon_T\rangle_V$.}
\label{colorfigures}
\end{center}
\end{figure}
Figure \ref{colorfigures} represents the isosurface plots of a snapshot of $T$ (top) and of $\epsilon_T$ (bottom) 
at $Ra=10^9$. We recognize the ridges in the temperature isosurface which correspond to the  plumes that 
are detached from the bottom and top boundary layers. Associated with the rising and descending plumes are 
larger amplitudes of the thermal dissipation field which can be observed in the bottom panel, similar to 
findings by Kerr (1996) or Shishkina \& Wagner (2007). The pattern of ridges (or stems) in the top panel 
is recaptured almost one-to-one in the bottom panel. The isosurface plot in the bottom panel indicates also 
that the local maxima of the thermal dissipation field $\epsilon_T$ are dominantly found close to both plates 
and not in the cell centre. 

In order to quantify this vertical dependence more precisely, we decompose the thermal dissipation rate field 
into contributions which result from the mean temperature profile and the temperature fluctuations. With 
(\ref{epsilont}), it follows
\begin{eqnarray}
\epsilon_T({\bf x},t)&=&\kappa \left[\left(\frac{\partial \langle T\rangle_A}{\partial z}\right)^2
                         +2\frac{\partial \langle T\rangle_A}{\partial z} 
                           \frac{\partial \theta}{\partial z} 
                         +({\bf\nabla}\theta)^2\right]\,,\\
                     &=&\epsilon_{\langle T\rangle}(z)+
                        2\kappa\frac{\partial \langle T\rangle_A}{\partial z} 
                               \frac{\partial \theta}{\partial z} +
                        \epsilon_{\theta}({\bf x},t)
\label{contributionstoeps}
\end{eqnarray}
The mixed term vanishes after averaging over planes at fixed height $z$ and thus 
\begin{equation}
\langle\epsilon_T\rangle_A(z)=\epsilon_{\langle T\rangle}(z)+
                            \langle\epsilon_{\theta}\rangle_A(z)\,.
\end{equation}
\begin{figure}
\centerline{\includegraphics[angle=0,scale=0.55,draft=false]{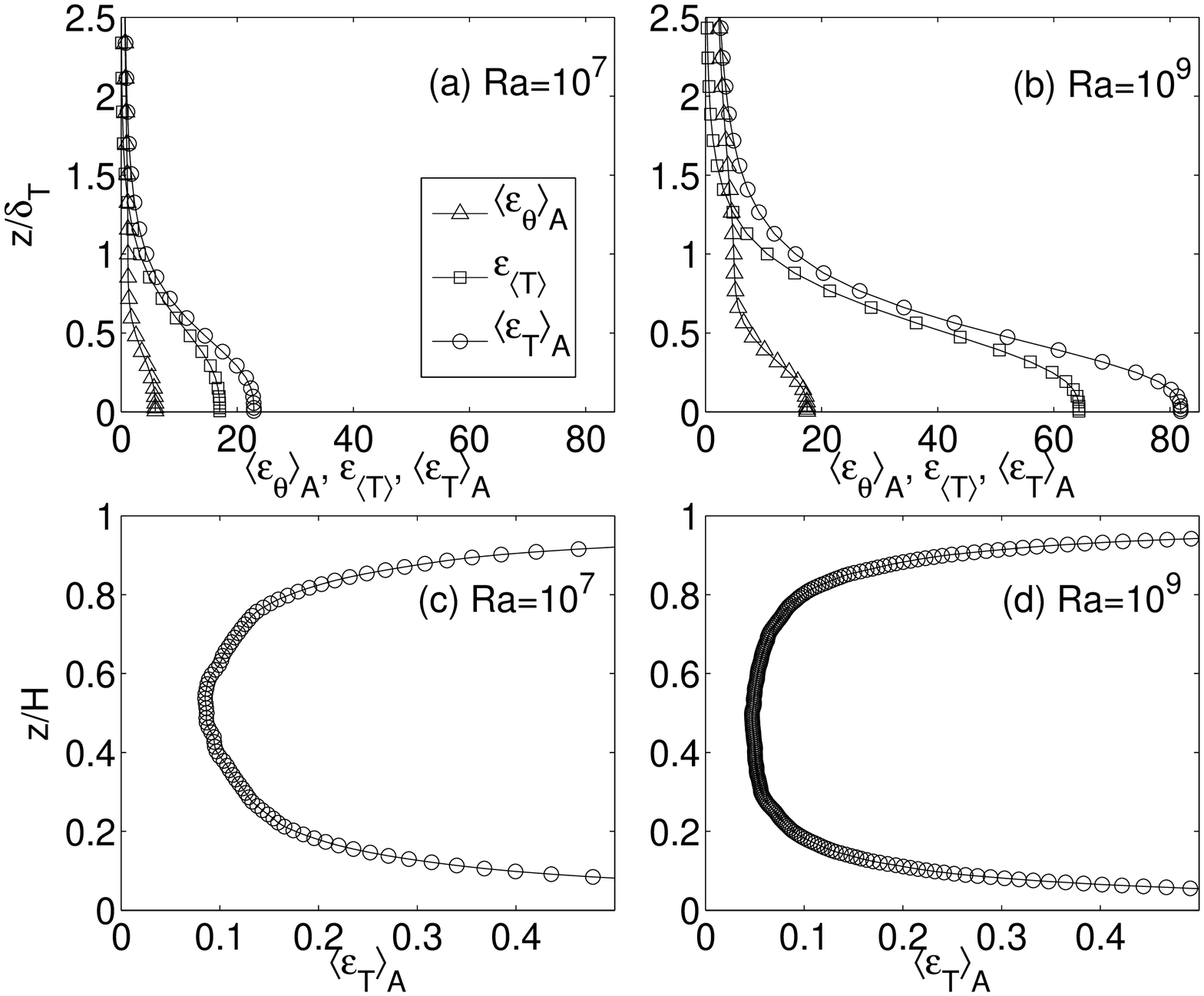}}
\caption{Vertical profiles of the two contributions to the total thermal dissipation 
rate which follow from the mean temperature and the temperature fluctuations 
(see (\ref{contributionstoeps})). Data are for Rayleigh number $Ra=10^7$ (a,c) and 
$Ra=10^9$ (b,d). The three different terms have been normalized by
the ensemble average of the total dissipation rate $\langle\epsilon_T\rangle_V=Nu/\sqrt{Ra Pr}$.
In panels (c) and (d), respectively, the vertical profiles of the total thermal dissipation
rate are replotted in order to demonstrate that they decrease significantly below the global ensemble
averages.}
\label{epsilonvertical}
\end{figure}
Figure \ref{epsilonvertical} plots the vertical profiles of both contributions to the total
thermal dissipation rate for Rayleigh numbers $Ra=10^7$ (left column) and $10^9$ (right column). 
All values of the thermal dissipation fields are given in units of the corresponding ensemble mean,
$Nu/\sqrt{Ra Pr}$ (see also (\ref{Nueps})). The contribution of the mean temperature profile,
$\epsilon_{\langle T\rangle}$, is dominant in the thermal boundary layer and decreases rapidly to 
almost zero toward the cell centre (see panels (a) and (b)). This mean profile contribution increases 
with increasing Rayleigh number since the scale across which the significant
temperature variation appears, $\delta_T$, becomes smaller. Conversely, the thermal dissipation rate 
due to the temperature fluctuations is dominant in the bulk and exceeds the mean profile contribution by 
more than 3 orders of magnitude there (not shown). As displayed 
in panels (c) and (d) of Fig. \ref{epsilonvertical} for both Rayleigh numbers, the total 
dissipation remains significantly below $Nu/\sqrt{Ra Pr}$ in the cell centre.  

From (\ref{Nueps}) and the scaling relation $Nu\sim Ra^{\beta}$, the Rayleigh-number dependence of the ensemble average $\langle\epsilon_T\rangle_V$ yields
\begin{equation}
\langle \epsilon_T\rangle_V\sim Ra^{\beta-\frac{1}{2}}=Ra^{\gamma}\,,
\end{equation}
which gives an exponent $\gamma=-0.217$ in our simulations. This global scaling can be refined 
for different subvolumes of the cylindrical domain. Well in the centre, we define the bulk dissipation as
\begin{equation}
\langle\epsilon_T\rangle_{Bulk}=\kappa\bigg\langle
\left(\frac{\partial T}{\partial x_i}\right)^2\bigg\rangle_{Bulk} 
\;\;\;\mbox{for}\;\;\;{\bf x}=(x,y,0.4\le z\le 0.6)\,. 
\label{bulkdiss}
\end{equation}
The thermal dissipation in the boundary layer is given by
\begin{equation}
\langle\epsilon_T\rangle_{BL}=\kappa\bigg\langle
\left(\frac{\partial T}{\partial x_i}\right)^2\bigg\rangle_{BL} 
\;\;\;\mbox{for}\;\;\;{\bf x}=(x,y,z\le \delta_T)\,. 
\label{bldiss}
\end{equation}
In line with the definitions of averages in section 2, we combine again an average over many 
realizations of the turbulent fields and a volume average which is taken now over the subvolumes $BL$ and
$Bulk$  as defined in (\ref{bulkdiss}) and (\ref{bldiss}), respectively.  Following Grossmann 
\& Lohse (2000) (see their eq. (2.15)), one can estimate the thermal dissipation in the
thermal boundary layer by assuming that it is due to the mean temperature drop across $\delta_T$ only and that
$\langle\epsilon_T\rangle_{BL}\approx\epsilon_{\langle T\rangle}$. This leads to
\begin{equation}
\langle\epsilon_T\rangle_{BL}\approx 
\kappa\left(\frac{\Delta T}{2\delta_T}\right)^2 \frac{\delta_T}{H} \sim 
\kappa\left(\frac{\Delta T}{H}\right)^2 Nu \sim Ra^{\beta}\,.
\label{GL1} 
\end{equation}
The factor $\delta_T/H$ stands for the fraction of the total volume which is occupied by the 
boundary layer. The last step in (\ref{GL1}) holds if $\kappa$ is a constant. For Rayleigh numbers between $5\times 10^7$ and $10^9$, our DNS give a scaling 
with respect to Rayleigh number of
\begin{equation}
\langle\epsilon_T\rangle_{BL}\approx 0.016\times Ra^{0.10}\,.
\end{equation}
This exponent is significantly smaller than the predictions by the scaling theories, 
$\beta=2/7$ or 1/3. We conclude that (\ref{GL1}) is not capturing the significant contribution 
from $\langle \epsilon_{\theta}\rangle_{BL}$ which can also be seen in Fig. \ref{epsilonvertical}. 
In addition, we find that $\langle\epsilon_{\theta}\rangle_{BL}$ displays almost the same scaling 
with Rayleigh number as $\langle\epsilon_T\rangle_{BL}$ in our direct numerical simulations.

In the bulk of the cylindrical cell, the thermal dissipation rate is due to temperature 
fluctuations only and can be estimated by (see eq. (2.12) in Grossmann \& Lohse (2000))
\begin{equation}
\langle\epsilon_T\rangle_{Bulk}\approx\langle\epsilon_{\theta}\rangle_{Bulk}\sim 
\frac{U (\Delta T)^2}{H}=\kappa \frac{(\Delta T)^2}{H^2} Re Pr\,,
\label{GL2} 
\end{equation}
Further explanation requires the knowledge of $Re(Ra, Pr)$. Corresponding to experiments, an anomalous 
scaling dependence of the Reynolds number on the Rayleigh number is found, i.e. deviations from a 
scaling exponent of $\zeta=0.5$ (Grossmann \& Lohse 2002; Brown {\it et al.} 2007a). Furthermore, 
the definition of the Reynolds number, especially the choice of the characteristic velocity, leads 
to additional variations in the exponents. In summary, the range of exponents is found to be between 
$\zeta=0.43$ and 0.49. It follows thus from (\ref{GL2}) that 
\begin{equation}
\langle\epsilon_T\rangle_{Bulk}\sim  Ra^{\zeta}\,,
\label{GL3} 
\end{equation}
for a fixed Prandtl number, as in the present study, with an exponent $\zeta>0$. However, our numerical simulations 
lead to
\begin{equation}
\langle\epsilon_T\rangle_{Bulk}\approx 0.32\times Ra^{-0.38}\,,
\label{scaling2}
\end{equation}
which comes close to the experimental findings by He {\it et al.} (2007) and exhibits the reported 
trend of Verzicco \& Camussi (2003). It does not vary significantly if the bulk volume fraction is 
increased in the analysis. For example, an exponent of $-0.34$ in (\ref{scaling2}) was obtained in the 
region $0.2\le z\le 0.8$. The Rayleigh number range, for which this least square fit was taken, is 
between $5\times 10^6$ and $10^9$. We conclude that the scaling exponent, which is by the way rather 
robust with respect to changes of the volume fraction, has the opposite sign in contrast to the 
scaling based on the estimate by Grossmann \& Lohse (2000). The reason  for this  discrepancy could be 
that the dimensionally correct ansatz $U (\Delta T)^2/H$ may not be appropriate for the thermal 
dissipation rate. Similar to fluid turbulence, the Rayleigh numbers (and therefore the Reynolds numbers) 
are still too small to reach the regime where the thermal dissipation rate given in units of $U (\Delta T)^2/H$ 
is a constant. This is known as the dissipation anomaly (Donzis, Sreenivasan \& Yeung, 2005). Both $\langle\epsilon_T\rangle_{Bulk}/(U_{rms}T^2_{rms}/H)$ and $\langle\epsilon_T\rangle_{Bulk}
/(U_{rms}\Delta T^2/H)$ still decrease with increasing $Ra$ for our data record ($U_{rms}$ defined as described
in the caption of table 1).  
\begin{figure}
\centerline{\includegraphics[angle=0,scale=0.5,draft=false]{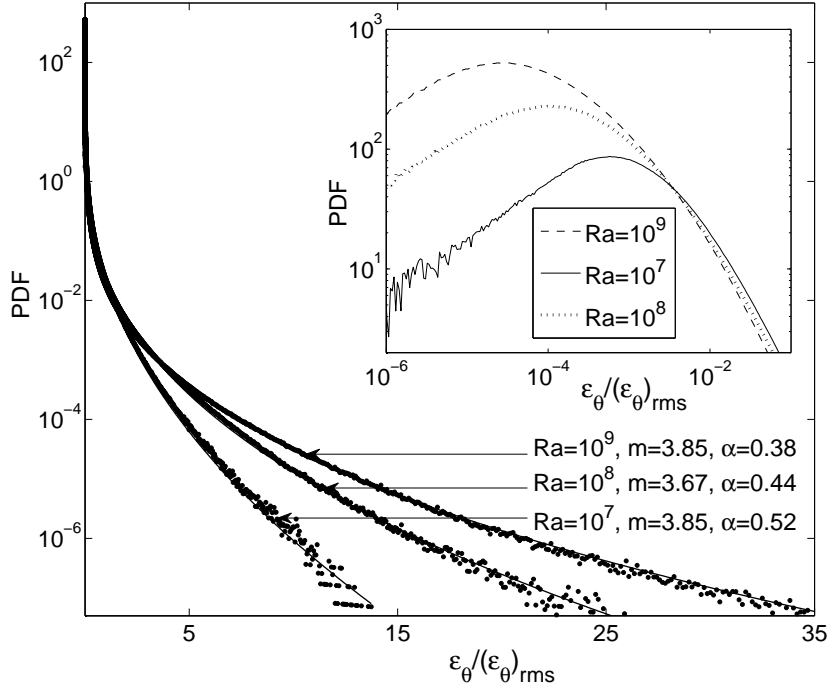}}
\caption{Tails of the PDFs of the thermal dissipation rates of the temperature fluctuations, $\epsilon_{\theta}$, in the bulk. 
Planes with $z\in [4\delta_T, H-4\delta_T]$ are included. The tails have been fitted by stretched 
exponentials as given by (\ref{stretched}) with the fit coefficients and exponents as indicated in the figure. The inset 
magnifies the PDFs for the smallest dissipation rate amplitudes in log-log scale. The data are normalized with respect to the rms value calculated over the whole volume and a sequence of statistically independent snapshots.}
\label{epsilonstretched}
\end{figure}
\begin{figure}
\centerline{\includegraphics[angle=0,scale=0.5,draft=false]{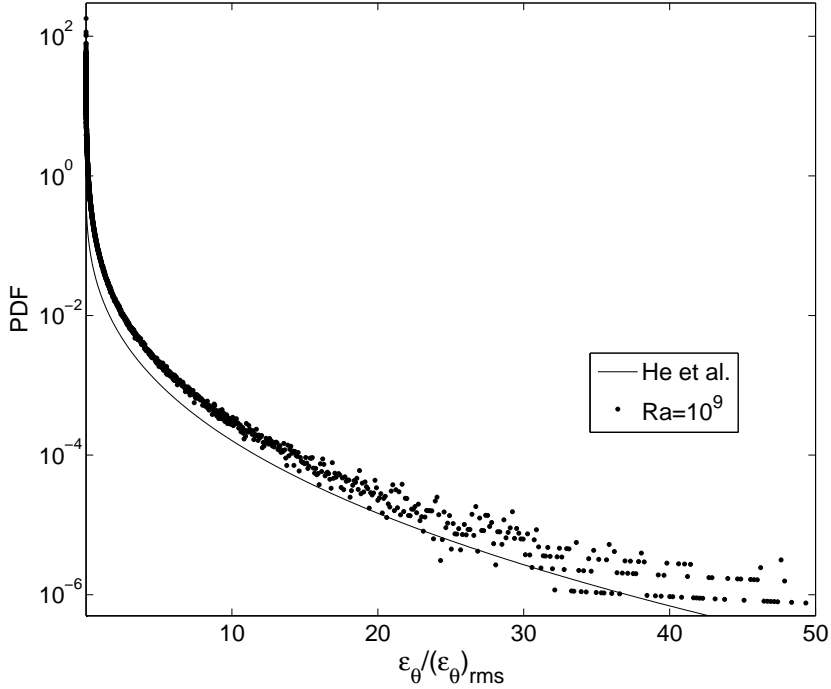}}
\caption{PDFs of $\epsilon_{\theta}$ at the cell centre (z=0.5H) normalized with respect to $(\epsilon_{\theta})_{rms}$, which is also taken at z=0.5H. Present DNS data at $Ra=10^9$ are compared with the experimental results (FIG. 4) of He {\it et al}. (2007) for $Ra=9.6\times 10^8$ and $Pr=5.4$. Statistics is gathered again over a sequence of independent snapshots. Data are more scattered in the far tail 
compared to Fig. \ref{epsilonstretched}.}
\label{compareHe}
\end{figure}
\begin{figure}
\centerline{\includegraphics[angle=0,scale=0.5,draft=false]{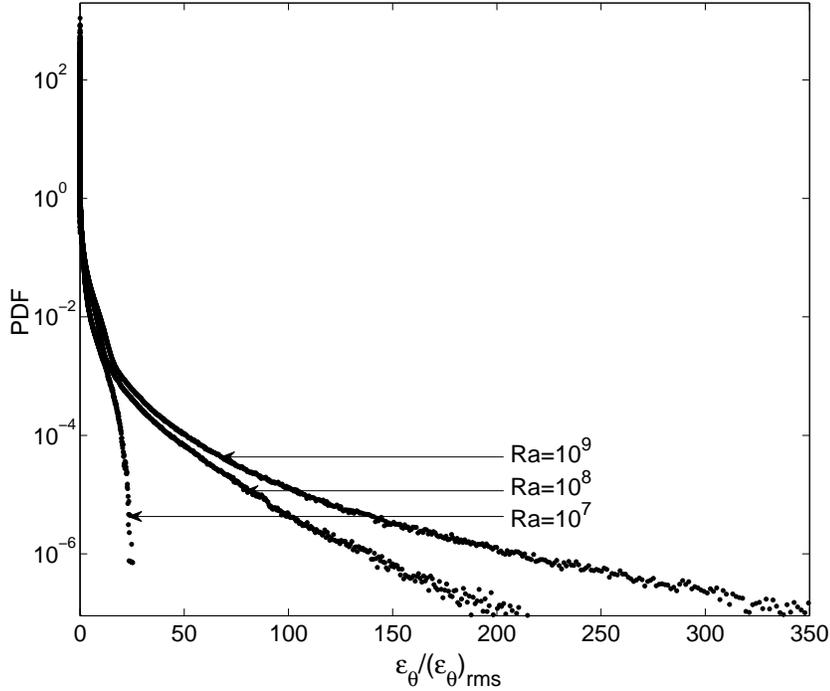}}
\caption{Tails of the PDFs of $\epsilon_{\theta}$ close to the bottom plate ($z\in [0, 4\delta_T]$) for different $Ra$. The data are normalized with respect to the rms value calculated over the whole volume and a sequence of statistically independent snapshots.}
\label{epsilonstretched1}
\end{figure}
\begin{figure}
\centerline{\includegraphics[angle=0,scale=0.6,draft=false]{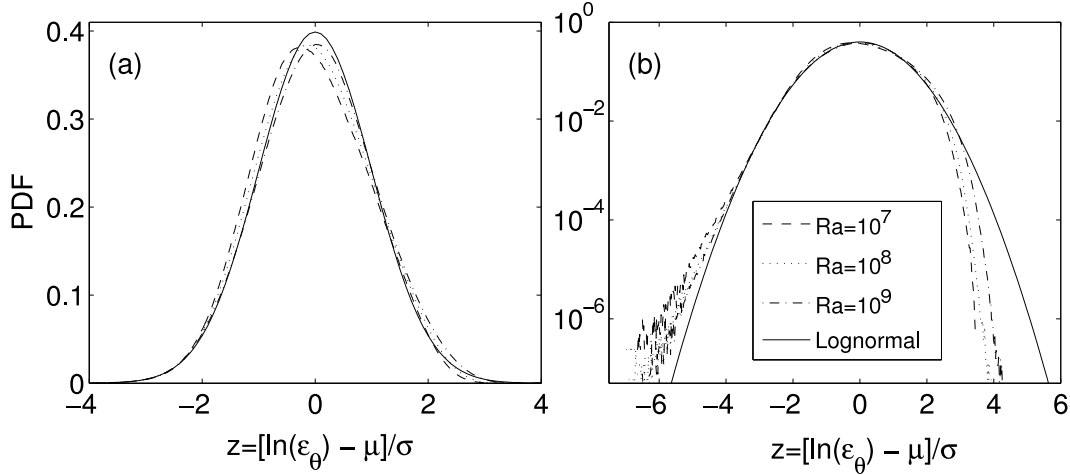}}
\caption{Replot of the same data as for Fig. \ref{epsilonstretched} in log-normal coordinates.
$\mu$ is the mean of $\ln(\epsilon_{\theta})$ and $\sigma$ is the corresponding standard deviation.
The solid line indicates a log-normal distribution. (a) Linear-linear plot. (b) Linear-logarithmic plot
of the same data. Rayleigh numbers are given in the legend.}
\label{epsilonlognormal}
\end{figure}
\begin{figure}
\centerline{\includegraphics[angle=0,scale=0.6,draft=false]{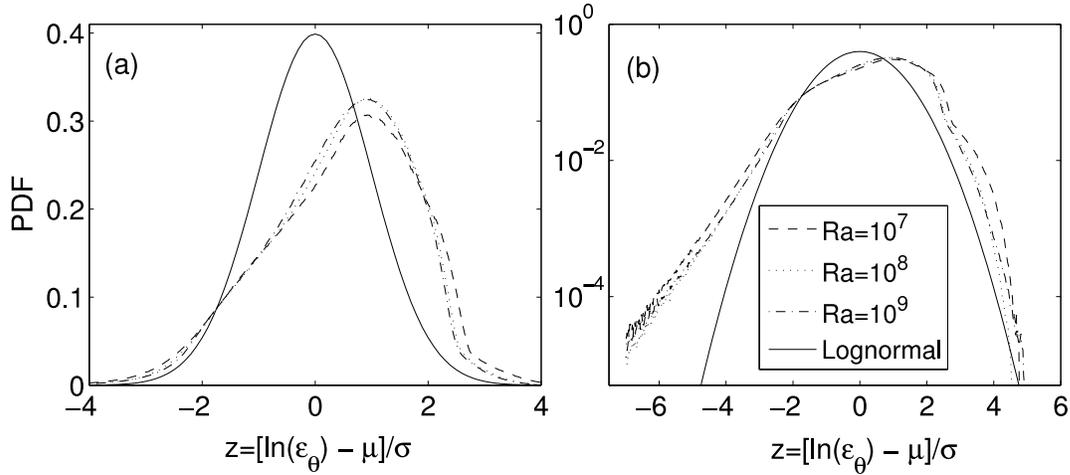}}
\caption{Replot of the same data as for Fig. \ref{epsilonstretched1} in log-normal coordinates.
Line and symbol styles are the same as in Fig. \ref{epsilonlognormal}.}
\label{epsilonlognormal1}
\end{figure}
\begin{figure}
\centerline{\includegraphics[angle=0,scale=0.5,draft=false]{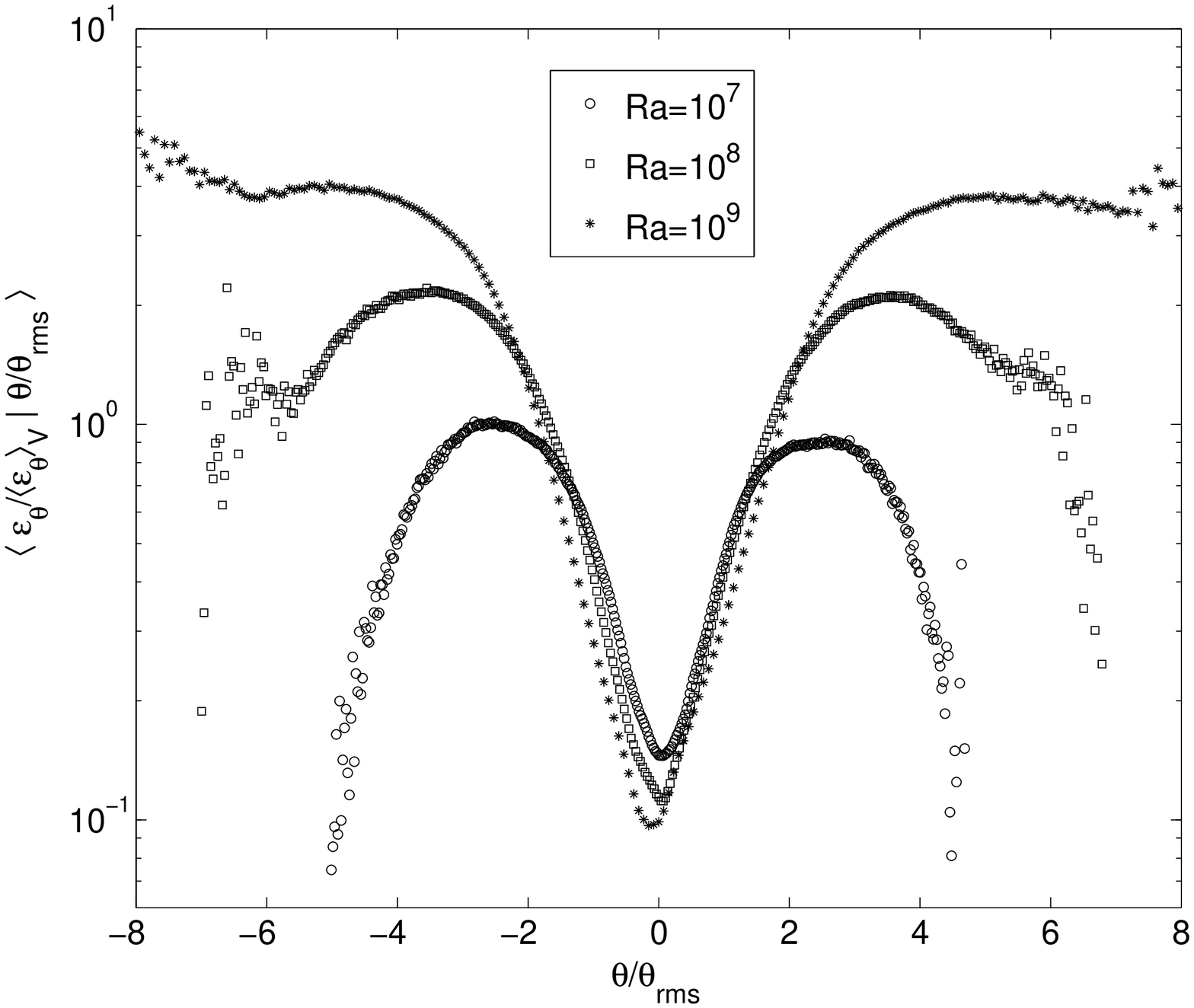}}
\caption{Conditional mean dissipation in the bulk (between $z\in [0, 4\delta_T]$ and $z \in [H-4\delta_T, H]$) as a function of the Rayleigh number. The rms of $\theta$ has been taken over the whole volume and a sequence of snapshots.}
\label{conditional}
\end{figure}

\subsection{Stretched exponential behaviour in the tails}
The mismatch of the Rayleigh number scaling with the predictions by Grossmann \& Lohse (2000) 
suggests that strong fluctuations of the thermal dissipation were not taken into account 
in the scaling. We discuss therefore the statistics of $\epsilon_{\theta}$ in this section. 
Figure \ref{epsilonstretched} shows the PDF of the dissipation rate of temperature fluctuations $\epsilon_{\theta}$ in 
the bulk. The analysis is conducted in planes between $z\in [4 \delta_T, H-4 \delta_T]$. In 
correspondence with the passive scalar case (Overholt \& Pope, 1996; Schumacher \& Sreenivasan, 2005), 
we fit a stretched exponential to the fraction of the PDF which extends from the most probable amplitude 
(see the inset of Fig. \ref{epsilonstretched}) to the end of the tail. We define $X=\epsilon_{\theta}/
\langle\epsilon_{\theta}\rangle_V$ and $X^{\ast}=X-X_{mp}$, with $X_{mp}$ is the abscissa of the most probable (mp) value. Data are fitted to
\begin{equation}
p(X^*)=\frac{C}{\sqrt{X^*}}\exp\left(-m X^{\ast\,\alpha}\right)\,.
\label{stretched}
\end{equation}
Similar to the passive scalar case, we do observe fatter tails with increasing Rayleigh number (and therefore 
with increasing Reynolds number). This is in line with an increasing degree of small-scale intermittency of 
the thermal dissipation rate field. Overholt \& Pope (1996) and Schumacher \& Sreenivasan (2005) have conducted 
a similar analysis for the passive scalar dissipation rate in a homogeneous isotropic turbulent flow. There, 
a minimal exponent $\alpha=1/3$  can be derived analytically in the limit of a very large Pecl\'{e}t number. 
The scalar is then advected in a flow which is white in time (Chertkov {\it et al.} 1998, Gamba \& Kolokolov 1999). 
The advection of a passive scalar in a Navier-Stokes flow always resulted in tails that were sparser, i.e. 
$\alpha\ge 1/3$ (Yeung {\it et al.} 2005). Our present analysis indicates that this threshold is also approached
for the hard-turbulence regime of convection.

Figure \ref{compareHe} replots our data in the center plane for $Ra=10^9$ in units of the rms value  of thermal dissipation
$(\epsilon_{\theta})_{rms}$ in order to compare them with the measurements by  He {\it et al.} (2007). A slightly sparser tail in the experiments is observed. On the one hand, the reason for this difference might be due to the improved resolution 
of the temperature gradient in our simulations, which was not possible in the experiment, as mentioned before. 
On the other hand, the Prandtl number ($Pr=5.4$) was almost an order of magnitude larger in the experiment. 
An increasing Prandtl number reduces the thermal boundary layer thickness and can cause less plume mixing  in the bulk. Less pronounced plumes will cause smaller amplitudes of the thermal
dissipation. Such a trend with the Prandtl number would differ from passive scalar turbulence, where the 
tails of the PDF of $\epsilon_{\theta}$ become fatter with increasing Prandtl (or Schmidt) number at a 
given Reynolds number of the advecting flow (Schumacher \& Sreenivasan, 2005). There a less diffusive
scalar field generates sharper gradients. It remains open which of those two effects is more dominant. 
Further simulations are required to study the dependence of the temperature gradient statistics on the Prandtl number in thermal convection.

When repeating the analysis for the part of the volume ($z\in [0, 4\delta_T]$) that has been excluded before, distributions as shown in Fig. \ref{epsilonstretched1} result. It is observed that the amplitudes of the dissipation 
increase by an order of magnitude for the largest $Ra$ compared to Fig. \ref{epsilonstretched}. The fatter tails 
for the boundary layer data are in line with the analysis of the vertical plane averaged profiles (see Fig. 
\ref{epsilonvertical}). The fluctuations of the thermal dissipation rate increase with increasing $Ra$.

\subsection{Deviations from log-normality}
The  intermittent nature of the thermal dissipation field leads to deviations from log-normality, as 
can be seen in Fig. \ref{epsilonlognormal} where the same  data set has been studied  as in 
Fig. \ref{epsilonstretched}. In order to highlight the differences in PDFs, we represent the data in linear-linear and linear-logarithmic axes respectively. It can be seen that neither
the core nor both tails of the PDF fit perfectly. This is similar to the passive scalar mixing. Small-amplitude 
tails are fatter than the log-normal curve while large-amplitude tails remain sparser (Ferchichi \& 
Tavoularis 2002, Schumacher \& Sreenivasan 2005). The trend of the data with Rayleigh numbers between 
$10^7$ and $10^9$ suggests a very slow convergence towards the log-normality. The repetition of the analysis in the vicinity of the thermal boundary 
layers, i.e. for the same data set as in Fig. \ref{epsilonstretched1}, reveals even stronger deviations 
(see Fig. \ref{epsilonlognormal1}). For the large-amplitude tail, a bump is observed. A similar feature 
has been reported by Kaczorowski \& Wagner (2007). Our analysis suggests that this particular 
feature is due to the boundary layer dynamics. The bump in the distribution remains also when the PDF 
is analysed for the whole cell and not for a particular subvolume.

\subsection{Conditional mean thermal dissipation}  
Deviations from the Gaussian distributions of the temperature fluctuation are also manifested in the conditional 
mean thermal dissipation. It is known that a conditional mean scalar dissipation is constant for a 
Gaussian scalar field, i.e. independent with respect to the scalar amplitude (see e.g. Overholt
\& Pope 1996). The mean is defined as 
\begin{equation}
\langle\epsilon_{\theta}|\theta({\bf x},t)=\psi\rangle=\int_0^{\infty}\mbox{d}\epsilon_{\theta}\,\epsilon_{\theta}\,
\frac{p(\epsilon_{\theta},\theta({\bf x},t)=\psi)}{p(\theta({\bf x},t)=\psi)}\,,       
\end{equation}
where $p(\epsilon_{\theta},\theta)$ is the joint PDF of the thermal dissipation rate and the temperature 
fluctuation. Figure \ref{conditional} displays clearly the V-shape for the conditional mean for temperature fluctuation values around $\theta/\theta_{rms}=0$. This is consistent 
with the deviations from Gaussianity, which were detected for the distributions of the temperature field. At 
larger magnitudes of $\theta$, the graphs reach a maximum before decreasing to zero since the support of the 
joint PDF and the temperature PDF is bounded. The V-shape is more pronounced with increasing Rayleigh number 
which is due to the higher thermal dissipation amplitudes observable in the flow (see Fig. \ref{epsilonstretched}). 
Our results agree with the findings of Jayesh \& Warhaft (1992).    

\section{Conclusions}
We have presented a detailed numerical study of the statistics of the temperature fluctuations and 
their gradients in turbulent convection. The resolution constraints limited the study to moderate Rayleigh numbers between $10^7$ and $10^9$. We compared the fine-scale structure 
of the active temperature field to that of the passive scalar field, far away from the boundaries. In contrast to the passive scalar case, the temperature statistics is {\em always} non-Gaussian with a probability density function close to exponential in the bulk. Non-Gaussianity holds in all regions of the convection cell and for all mentioned Rayleigh 
numbers. Deviations in temperature fluctuations from the Gaussian distribution have also been  confirmed by the V-shaped conditional 
mean dissipation. For the passive scalar fluctuations, the statistics depends on the mechanism 
that sustains the fluctuations and on the particular ratios of outer turbulence length scales. Super-Gaussian, Gausisan or weakly sub-Gaussian distributions have been observed there in the past. While the 
statistics of the temperature can differ from the passive scalar, the statistics of the spatial derivatives 
and dissipation rates behave qualitatively similarly.

The probability density function of the thermal dissipation rate in the bulk
is fitted well with a stretched exponential. The tails extend to larger amplitudes with increasing Rayleigh
number. This is a clear fingerprint of a stronger small-scale intermittency. Thermal dissipation field is 
always more intermittent in the boundary layer than the bulk of the convection cell. This is indicated by the extended tail of the PDF. All distributions showed clear deviations from the log-normality in the whole range of dissipation values. 

Motivated by similar studies in passive scalar turbulence and by our detection of locally varying mean temperature profiles, the deviations from locally isotropic temperature 
fluctuations are quantified by the third order vertical derivative moments as a function of the Rayleigh number. 
Interestingly, we found that the deviations from the local isotropy grow with increasing $Ra$ in the boundary layer. In the bulk, the derivative skewness is found to decrease. Our data exhibit return to isotropy trend (Lumley 1967) when translating the original Reynolds number dependence of the skewness into a Rayleigh number dependence for the convection case. This is in contrast to the passive scalar case which does not show such trend (Warhaft 2000). The observed return might be due to the small amplitudes that the local mean temperature gradients possess. The vertical profiles of the plane averaged thermal dissipation show furthermore that an increasing fraction 
is concentrated in an ever thinner boundary layer with growing $Ra$. Finally, we demonstrate that the contribution of the temperature fluctuations to the total thermal dissipation in the thermal boundary layer is significant and cannot be neglected. This aspect is not included in the scaling  of the thermal dissipation rate with 
respect to the Rayleigh number and can explain why the present trends of the total thermal dissipation with 
respect to the Rayleigh number differ in comparison to the scaling theory by Grossmann \& Lohse (2000).

Two aspects will be addressed in our future work on this subject. Firstly, an extension to larger aspect 
ratios and higher Rayleigh numbers is desirable. It is known from experiments, such as by du Puits {\it et al.} (2007a), that the large-scale motion is sensitive to the aspect ratio. A rearrangement of the large-scale flow patterns will consequently affect the  statistics of various quantities. Secondly, the direct 
link between the observed statistics and the local structures in the boundary layer and its vicinity is necessary. 
First steps have been done already in Shishkina \& Wagner (2007) and Zhou {\it et al.} (2007). Our investigations 
on both aspects have started recently and will be discussed elsewhere.
 
\begin{acknowledgments}
We would like to thank first Roberto Verzicco for providing us his simulation code and his help 
at the beginning of our studies. We acknowledge discussions with Norbert Peters, Katepalli R. Sreenivasan 
and Andr\'{e} Thess. We also thank for supercomputing ressources on the IBM p690 cluster JUMP of the 
J\"ulich Supercomputing Centre (Germany) under grant HMR09. The work is supported by the Deutsche 
Forschungsgemeinschaft (DFG) under grant SCHU1410/2 and by the Heisenberg Program of the DFG. We also wish 
to thank all three referees for their constructive comments and suggestions.
\end{acknowledgments}


\begin{thebibliography}{99}

\bibitem[Belmonte (1994)]{Belmonte1994}
{\sc Belmonte A., Tilgner, A. \& Libchaber, A.} 1994
Temperature and velocity boundary layers in turbulent convection.
{\it Phys. Rev. E} {\bf 50}, 269-279.

\bibitem[Belmonte (1996)]{Belmonte1996}
{\sc Belmonte A. \& Libchaber, A.} 1996
Thermal signature of plumes in turbulent convection: The skewness of the derivative.
{\it Phys. Rev. E} {\bf 53}, 4893-4898.

\bibitem[Brown (2007)]{Brown2007}
{\sc Brown, E. \& Ahlers, G.} 2007
Temperature gradients, and search for non-Boussinesq effects, in the interior of turbulent Rayleigh-B\'{e}nard convection.
{\it Europhys. Lett.} {\bf 80}, 14001 (6 pages).

\bibitem[Brown (2007a)]{Brown2007a}
{\sc Brown, E., Funfschilling, D. \& Ahlers, G.} 2007a
Anomalous Reynolds-number scaling in turbulent Rayleigh-B\'{e}nard convection.
{\it J. Stat. Mech.} P10005.

\bibitem[Castaing (1989)]{Castaing1989}
{\sc Castaing, B., Gunaratne, G., Heslot, F., Kadanoff, L., Libchaber, A., Thomae, S., Wu, X.-Z.,
Zaleski, S. \& Zanetti, G. } 1989
Scaling of hard thermal turbulence in Rayleigh-B\'{e}nard convection.
{\it J. Fluid Mech.} {\bf 204}, 1-30.

\bibitem[Castaing (1990)]{Castaing1990}
{\sc Castaing, B., Gagne, Y. \& Hopfinger, E. J.} 1990
Velocity probability density functions of high Reynolds number turbulence.
{\it Physica D} {\bf 46}, 177-200.
 
\bibitem[Chertkov (1998)]{Chertkov1998}
{\sc Chertkov, M., Falkovich, G. \& Kolokolov, I.} 1998  
Intermittent dissipation of a passive scalar in turbulence.
{\it Phys. Rev. Lett.} {\bf 80}, 2121-2124.

\bibitem[Chevillard (2006)]{Chevillard2006}
{\sc Chevillard, L., Castaing, B., L\'{e}v\^{e}que \& Arneodo, A.} 2006
Unified multifractal description of velocity increments statistics in turbulence: Intermittency and skewness.
{\it Physica D} {\bf 218}, 77-82.

\bibitem[Ching (1991)]{Ching1991}
{\sc Ching, E. S. C.} 1991
Probabilities for temperature differences in Rayleigh-B\'{e}nard convection.
{\it Phys. Rev. A} {\bf 44}, 3622-3628.

\bibitem[Ching (1993)]{Ching1993}
{\sc Ching, E. S. C.} 1993
Probability densities of turbulent temperature fluctuations.
{\it Phys. Rev. Lett.} {\bf 70}, 283-286.

\bibitem[Dimotakis (2005)]{Dimotakis2005} 
{\sc Dimotakis, P. E.} 2005
Turbulent mixing.
{\it Annu. Rev. Fluid Mech.}  {\bf 37} 329-356.

\bibitem[Donzis (2005)]{Donzis2005}
{\sc Donzis, D. A., Sreenivasan, K. R. \& Yeung, P. K. } 2005
Scalar dissipation rate and dissipative anomaly in isotropic turbulence.
{\it J. Fluid Mech.} {\bf 532}, 199-216.

\bibitem[du Puits (2007)]{duPuits2007}
{\sc du Puits, R., Resagk, C., Tilgner, A., Busse, F.-H. \& Thess, A.} 2007
Structure of thermal boundary layers in turbulent Rayleigh-B\'{e}nard convection.
{\it J. Fluid Mech.} {\bf 572}, 231-254.

\bibitem[du Puits (2007a)]{duPuits2007a}
{\sc du Puits, R., Resagk, C. \& Thess, A.} 2007a
Breakdown of wind in turbulent thermal convection.
{\it Phys. Rev. E} {\bf 75}, 016302 (4 pages).

\bibitem[Ferchichi (2002)]{Ferchichi2002}
{\sc Ferchichi, M. \& Tavoularis, S.} 2002
Scalar probability density function and fine structure in uniformly sheared turbulence.
{\it J. Fluid Mech.} {\bf 461}, 155-182.

\bibitem[Funfschilling (2005)]{Funfschilling2005}
{\sc Funfschilling, D., Brown, E., Nikolaenko, A. \& Ahlers, G.} 2005
Heat transport by turbulent Rayleigh-B\'{e}nard convection in cylindrical samples with aspect
ratio one and larger.
{\it J. Fluid Mech.} {\bf 536}, 145-154.

\bibitem[Gamba (1999)]{Gamba1999}
{\sc Gamba A. \& Kolokolov, I.} 1999
Dissipation statistics of a passive scalar in a multi-dimensional smooth flow.
{\it J. Stat. Phys.} {\bf 94}, 759-777.

\bibitem[Gollub (1991)]{Gollub1991}
{\sc Gollub, J. P., Clarke, J., Gharib, M., Lane, B. \& Mesquita, O. N.} 1991
Fluctuations and transport in a stirred fluid with a mean gradient.
{\it Phys. Rev. Lett.} {\bf 67}, 3507-3510.

\bibitem[Groetzbach (1983)]{Groetzbach1983}
{\sc Gr\"otzbach, G.} 1983
Spatial resolution requirements for direct numerical simulation of the Rayleigh-B\'{e}nard
convection.
{\it J. Comput. Phys.} {\bf 49}, 241-269.

\bibitem[Grossmann (2000)]{Grossmann2000}
{\sc Grossmann, S. \& Lohse, D.} 2000
Scaling in thermal convection: a unifying theory.
{\it J. Fluid Mech.} {\bf 407}, 27-56.

\bibitem[Grossmann (2002)]{Grossmann2002}
{\sc Grossmann, S. \& Lohse, D.} 2002
Prandtl and Rayleigh number dependence of the Reynolds number in turbulent thermal convection.
{\it Phys. Rev. E} {\bf 66}, 016305 (6 pages).

\bibitem[Gylfason (2004)]{Gylfason2004}
{\sc Gylfason, A. \& Warhaft, Z.} 2004
On higher order passive scalar structure functions in grid turbulence.
{\it Phys. Fluids} {\bf 16}, 4012-4019.

\bibitem[Hartlep (2005)]{Hartlep2005}
{\sc Hartlep, T., Tilgner, A. \& Busse, F. H.} 2005
Transition to turbulent convection in a fluid layer heated from below at moderate aspect ratio.
{\it J. Fluid Mech.} {\bf 544}, 309-322.
 
\bibitem[He (2007)]{He2007}
{\sc He, X., Tong, P. \& Xia, K.-Q.} 2007
Measured thermal dissipation field in turbulent Rayleigh-B\'{e}nard convection.
{\it Phys. Rev. Lett.} {\bf 98}, 144501 (4 pages).

\bibitem[Heslot (1987)]{Heslot1987}
{\sc Heslot, F., Castaing, B. \& Libchaber, A.} 1987
Transitions to turbulence in helium gas.
{\it Phys. Rev. A} {\bf 36}, 5870-5873.

\bibitem[Jayesh (1991)]{Jayesh1991}
{\sc Jayesh \& Warhaft, Z.} 1991
Probability distribution of a passive scalar in grid-generated turbulence.
{\it Phys. Rev. Lett.} {\bf 67}, 3503-3506.

\bibitem[Jayesh (1992)]{Jayesh1992}
{\sc Jayesh \& Warhaft, Z.} 1992
Probability distribution, conditional dissipation and transport of passive temperature fluctuations 
in grid-generated turbulence.
{\it Phys. Fluids A} {\bf 4}, 2292-2307.


\bibitem[Kadanoff (2001)]{Kadanoff2001}
{\sc Kadanoff, L. P.} 2001
Turbulent heat flow: Structures and scaling.
{\it Phys. Today} {\bf 54}, 34-99.

\bibitem[Kaczorowski (2007)]{Kaczorowski2007}
{\sc Kaczorowski, M. \& Wagner C.} 2007
Direct numerical simulation of turbulent convection in a rectangular Rayleigh-B\'{e}nard cell.
{\it Proceedings of Fifth International Symposium on Turbulence and Shear Flow Phenomena}, Garching, 2007, Vol.2, 499-504.

\bibitem[Kerr (1996)]{Kerr1996}
{\sc Kerr, R. M.} 1996
Rayleigh number scaling in numerical convection.
{\it J. Fluid Mech.} {\bf 310}, 139-179.

\bibitem[Kushnir (2006)]{Kushnir2006}
{\sc Kushnir, D., Schumacher, J. \& Brandt, A.} 2006
Geometry of intensive scalar dissipation events in turbulence.
{\it Phys. Rev. Lett.} {\bf 97}, 124502.

\bibitem[Lumley (1967)]{Lumley1967}
{\sc Lumley, J. L.} 1967
Similarity and the turbulent energy spectrum. 
{\it Phys. Fluids} {\bf 10}, 855-858. 

\bibitem[Mydlarski (1998)]{Mydlarski1998}
{\sc Mydlarski, L. \& Warhaft, Z.} 1998
Passive scalar statistics in high-P\'{e}clet-number grid turbulence.
{\it J. Fluid Mech.} {\bf 358}, 135-175.

\bibitem[Niemela (2000)]{Niemela2000}
{\sc Niemela, J. J., Skrbek, L., Sreenivasan, K. R. \& Donelly, R. J.} 2000
Turbulent convection at very high Rayleigh numbers.
{\it Nature} {\bf 404}, 837-840.

\bibitem[Niemela (2003)]{Niemela2003}
{\sc Niemela, J. J. \& Sreenivasan, K. R.} 2003
Turbulent confined convection.
{\it J. Fluid Mech.} {\bf 481}, 355-384.

\bibitem[Overholt (1996)]{Overholt1996}
{\sc Overholt, M. R. \& Pope, S. B.} 1996
Direct numerical simulation of a passive scalar with imposed mean gradient in isotropic turbulence.
{\it Phys. Fluids} {\bf 8}, 3128-3148.

\bibitem[Pumir (1996)]{Pumir1996}
{\sc Pumir, A.} 1996
Turbulence in homogeneous shear flows.
{\it Phys. Fluids} {\bf 8}, 3112-3127.

\bibitem[Pumir (1991)]{Pumir1991}
{\sc Pumir, A., Shraiman, B. \& Siggia, E. D.} 1991
Exponential tails and random advection.
{\it Phys. Rev. Lett.} {\bf 23}, 2984-2987.

\bibitem[Schumacher (2003)]{Schumacher2003}
{\sc Schumacher, J. \& Sreenivasan, K. R.} 2003
Geometric features of the mixing of passive scalars at high Schmidt numbers.
{\it Phys. Rev. Lett.} {\bf 91}, 174501 (4 pages).

\bibitem[Schumacher (2005)]{Schumacher2005}
{\sc Schumacher, J., Sreenivasan, K. R. \& Yeung, P. K.} 2005
Very fine structures in scalar mixing.
{\it J. Fluid Mech.} {\bf 531}, 113-122.

\bibitem[Schumacher (2005)]{Schumacher2005a}
{\sc Schumacher, J.\& Sreenivasan, K. R.} 2005
Statistics and geometry of passive scalars in turbulence.
{\it Phys. Fluids} {\bf 17}, 125107 (9 pages).

\bibitem[Schumacher (2008)]{Schumacher2008}
{\sc Schumacher, J.} 2008
Lagrangian dispersion and heat transport in convective turbulence.
{\it Phys. Rev. Lett.} {\bf 100}, 134502 (4 pages).

\bibitem[Shishkina (2006)]{Shishkina2006}
{\sc Shishkina, O. \& Wagner C.} 2006
Analysis of thermal dissipation rates in turbulent Rayleigh-B\'{e}nard convection. 
{\it J. Fluid Mech.} {\bf 546}, 51-60.

\bibitem[Shishkina (2007)]{Shishkina2007}
{\sc Shishkina, O. \& Wagner C.} 2007
Local heat flux in turbulent Rayleigh-B\'{e}nard convection. 
{\it Phys. Fluids} {\bf 19}, 085107 (13 pages).

\bibitem[Shraiman (2000)]{Shraiman2000}
{\sc Shraiman, B. I. \& Siggia E. D.} 2000
Scalar turbulence. 
{\it Nature} {\bf 405}, 639-646.

\bibitem[Siggia (1994)]{Siggia1994} 
{\sc Siggia E. D.} 1994
High Rayleigh number convection.
{\it Annu. Rev. Fluid Mech.}  {\bf 26} 137-68.
 
\bibitem[Verzicco (2003)]{Verzicco2003}
{\sc Verzicco, R. \& Camussi, R.} 2003
Numerical experiments on strongly turbulent thermal convection in a slender 
cylindrical cell.
{\it J. Fluid Mech.} {\bf 477}, 19-49.

\bibitem[Verzicco (1996)]{Verzicco1996}
{\sc Verzicco, R. \& Orlandi, P.} 1996
A finite-difference scheme for three-dimensional incompressible flows in 
cylindrical coordinates.
{\it J. Comp. Phys.} {\bf 123}, 402-414.

\bibitem[Warhaft (2000)]{Warhaft2000}
{\sc Warhaft, Z.} 2000
Passive scalars in turbulent flows.
{\it Annu. Rev. Fluid Mech.} {\bf 32}, 203-240.

\bibitem[Warhaft (2002)]{Warhaft2002}
{\sc Warhaft, Z.} 2002
Turbulence in nature and laboratory.
{\it Proc. Nat. Acad. Sci.} {\bf 99}, 2481-2486.

\bibitem[Watanabe (2004)]{Watanabe2004}
{\sc Watanabe, T. \& Gotoh, T.} 2004
Statistics of a passive scalar in homogeneous turbulence.
{\it New J. Phys.} {\bf 6}, 40 (36 pages).

\bibitem[Xi (2004)]{Xi2004}
{\sc Xi, H.-D., Lam, S. \& Xia, K.-Q.} 2004
From laminar plumes to organized flows: the onset of large-scale circulation in turbulent
thermal convection.
{\it J. Fluid Mech.} {\bf 503}, 47-56.

\bibitem[Yakhot (1989)]{Yakhot1989}
{\sc Yakhot, V.} 1989
Probability distributions in high-Rayleigh-number B\'{e}nard convection.
{\it Phys. Rev. Lett.} {\bf 63} 1965-1967. 

\bibitem[Yeung (2005)]{Yeung2005}
{\sc Yeung, P. K., Donzis, D. A. \& Sreenivasan, K. R.} 2005
High-Reynolds-number simulation of turbulent mixing.
{\it Phys. Fluids} {\bf 17} 081703 (4 pages). 

\bibitem[Zhou (2002)]{Zhou2002}
{\sc Zhou S.-Q. \& Xia, K.-Q.} 2002
Plume statistics in thermal turbulence: Mixing of an active scalar.
{\it Phys. Rev. Lett.} {\bf 89}, 184502 (4 pages).

\bibitem[Zhou (2007)]{Zhou2007}
{\sc Zhou Q., Sun, C. \& Xia, K.-Q.} 2007
Morphological evolution of thermal plumes in turbulent Rayleigh-B\'{e}nard convection.
{\it Phys. Rev. Lett.} {\bf 98}, 074501 (4 pages).

\end{thebibliography}
\end{document}